\documentclass[a4paper,8pt]{article}
\usepackage{natbib}
\usepackage{epsfig}
\usepackage{amssymb}
\usepackage{amsmath}
\usepackage{subfigure} 
\usepackage{psfrag}
\usepackage{graphicx}
\usepackage{gensymb}
\usepackage{float}
\usepackage{array}
\usepackage{mathrsfs}
\usepackage{fancyheadings}

\begin{document}
  \title{\bf \LARGE{Fluid kinematics around two circular cylinders moving towards impact}}
  \author{ \textsc{\Large{Nektarios Bampalas}}
	\footnote{Current email address for
      correspondence: nektbam@gmail.com}
	\\
  Department of Aeronautics, Imperial College London, United Kingdom}
\date{September 2009}
\maketitle
\begin{abstract}
 
  The scale factors of an arbitrary orthogonal space are a
  measure of its content of homogeneous orthogonal space. In
  the present study, it is shown, that their spatial and
  temporal rates of variation do not contribute to the
  differential calculus of arbitrary functions in orthogonal
  space. Based on this, the Navier-Stokes equations are
  formulated accordingly, to provide a method of studying the
  kinematics of fluid motion in orthogonal plane
  space. Employing this formulation and regular perturbation
  theory, the kinematic physical measures of the flow of an
  incompressible, viscous fluid around two identical circular
  cylinders, which move with equal and opposite velocity
  towards central impact are evaluated. For this case, the
  space is parametrised according to the bipolar
  transformation of the cartesian plane.
\end{abstract}
\section{Introduction}
\label{Section_1}

\hspace{3mm}
The expressions of the basic differential operators,
according to an arbitrary orthogonal vector base, are known,
see e.g. \cite{Spain}. The scale factors, which characterise
such a base, are, in general, functions of space and
time, and they represent a spatial density measure between
the two mapping spaces. The present study shows that their spatial
rate of variation does not contribute to the spatial rate of variation of arbitrary
functions of the orthogonal space.
	 
	 \hspace{3mm}
	 Based on this result, which affects the differential calculus
	 of functions in orthogonal space, the Navier-Stokes equations
	 are formulated in an arbitrarily parametrised, orthogonal,
	 plane space. An application of this formulation follows, in
	 order to obtain an asymptotic series solution for the
	 kinematic physical measures of the flow of a fluid around two
	 identical circular cylinders, whch move with equal and
	 opposite velocity, towards central impact.
	 
	 \hspace{3mm}
	 \cite{Bampalas_Graham} studied the flow, based on a numerical
	 solution of the Navier-Stokes equations in the
	 plane. However, as the boundaries of the cylinders approach,
	 mesh resolution effects affect the accuracy of the numerical
	 solution.
	 
	 \hspace{3mm}
	 The next section presents the expressions of the basic
	 differential operators in orthogonal space, on which the
	 subsequent analysis is based on. In section \ref{Section_3},
	 the formulation of the Navier-Stokes equations in an
	 arbitrarily parametrised, orthgonal, plane space is
	 presented. In section \ref{Section_4}, the bipolar
	 transformation of the cartesian plane is employed, to adapt
	 the formulation to the case of the flow of a fluid around two
	 circular cylinders and the asymptotic series solution is obtained.
	 
	 \section{The basic differential operators in an arbitrary
	   orthogonal space}
	 \label{Section_2}
	 
	 \hspace{3mm}
	 Consider the right-handed, orthogonal vector base
	 $\mathbf{I}_i$, $i\in \mathbb{N} : i \in [1,3]$, the spaces
	 $\mathbf{q}$, $\mathbf{s}\in\mathbb{R}^3$ and time $t\in
	 \mathbb{R}: t\ge 0$. Implying the summation convention,
	 \begin{align*}
	   \mathbf{q}:=q_i ~ \mathbf{I}_i \qquad, \qquad \mathbf{s}(\mathbf{q};t):=s_i(\mathbf{q};t)~\mathbf{I}_i~;
	 \end{align*}
	 $s_i(\mathbf{q};t)$ are independent variables in $\mathbf{s}$
	 and $q_i$ are independent variables in $\mathbf{q}$.
	 
	 \hspace{3mm}
	 The units of spatial measure of $\mathbf{q}$ and $\mathbf{s}$
	 are
	 
	 \begin{align*}
	   d\mathbf{q}:=\{dq_i \mathbf{I}_i \in \mathbf{q} : dq_i \to
	   0\} \neq 0 \quad , \quad d\mathbf{s}:=\{ ds_i
	   \mathbf{I}_i \in\mathbf{s}:ds_i\to 0\} \neq 0 \quad.
	 \end{align*}
	 The variation of arbitrary, continuous functions of
	 $\mathbf{q}$ or $\mathbf{s}$ along $d\mathbf{q}$ or
	 $d\mathbf{s}$ is assumed to be linear.
	 
	 \hspace{3mm}
	 Define a mapping
	 \begin{align*}
	   \mathscr{M}:=\{\mathbf{s} \mapsto \mathbf{q}~:
	   ~ds_i\mathbf{I}_i=\mathfrak{s}_i(\mathbf{q};t)dq_i\mathbf{I}_i ~ ,
	   ~
	   \mathfrak{s}_i(\mathbf{q};t)\in\mathbb{R}:\mathfrak{s}_i\neq0
	   \} ~ .
	 \end{align*}
	 $\mathfrak{S}(\mathbf{q};t):=\mathfrak{s}_i\mathbf{I}_i$
	 represents a spatial density measure of $\mathbf{q}$ in $\mathbf{s}$. For
	 the special case $\mathbf{q} ,\mathbf{s}\in \mathbb{R}^2$ and
	 for the bipolar transformation, $\mathfrak{s}_i=\mathfrak{s}$.
	 \hspace{3mm}
	 $\mathbf{s}$ is measured as
	 \begin{align*}
	   \mathbf{s}=\int_0^{s_i}du~\mathbf{I}_i=\int_0^{q_i}\mathfrak{s}_i[\mathbf{q}(u);t]du~\mathbf{I}_i~.
	 \end{align*}
	 
	 \hspace{3mm}
	 Consider an arbitrary, continuous scalar function
	 $f(\mathbf{s},t)\in\mathbb{R}$, with continuous
	 derivatives. The spatial rate of variation of $f$ along
	 $\mathbf{I}_i$ is described by the directional derivative
	 $f_{s_i}$ along $\mathbf{I}_i$ (see e.g. \cite{Kreyszig}),
	 which is defined as
	 \begin{align*}
	   f_{s_i}[\mathbf{s}(\mathbf{q};t),t]&=\frac{df}{ds_i}:=\lim_{ds_i\to
	     0}\frac{f[\mathbf{s}(\mathbf{q};t)+ds_i(\mathbf{q};t)\mathbf{I}_i,t]-f[\mathbf{s}(\mathbf{q};t),t]}{ds_i}=\\
	   &=\lim_{\mathfrak{s}_i(\mathbf{q};t)dq_i\to
	     0}\frac{f[\mathbf{s}(\mathbf{q};t)+\mathfrak{s}_i(\mathbf{q};t)dq_i\mathbf{I}_i,t]-f[\mathbf{s}(\mathbf{q};t),t]}{\mathfrak{s}_i(\mathbf{q};t)dq_i}=\\
	   &=\mathfrak{s}_i^{-1}(\mathbf{q};t)\lim_{dq_i\to 0}\frac{f[\mathbf{q}+dq_i\mathbf{I}_i,t;\mathfrak{S}(\mathbf{q};t)]-f[\mathbf{q},t;\mathfrak{S}(\mathbf{q};t)]}{dq_i}~.
	 \end{align*}
	 Therefore,
	 
	 \begin{align*}
	   f_{s_i}[\mathbf{s}(\mathbf{q};t),t]=f_{s_i}[\mathbf{q},t;\mathfrak{S}(\mathbf{q};t)]=\mathfrak{s}_i^{-1}(\mathbf{q};t)f_{q_i}[\mathbf{q},t;\mathfrak{S}(\mathbf{q};t)]~.
	 \end{align*}
	 
	 \hspace{3mm}From the first fundamental theorem of calculus (see
	 e.g. \cite{Apostol}), the indefinite integral of $f_{s_i}$
	 along $\mathbf{I}_i$, is
	 
	 \begin{align*}
	   f[\mathbf{s}(\mathbf{q};t),t]=f[\mathbf{q},t;\mathfrak{S}(\mathbf{q};t)]=\int_{\xi}^{s_i}f_{s_i}[\mathbf{s}(u),t]du=
	   \int_{\zeta}^{q_i}f_{q_i}[\mathbf{q}(u),t;\mathfrak{S}(\mathbf{q};t)]du~,
	 \end{align*}
	 where $\xi\in\mathbf{s}\cdot \mathbf{I}_i$,
	 $\zeta\in\mathbf{q}\cdot \mathbf{I}_i$.
	 
	 \hspace{3mm}
	 Space and time are considered to be independent
	 sets. Therefore, the temporal rate of variation of
	 $f[\mathbf{s}(\mathbf{q};t),t]$ is
	 
	 \begin{align*}
	   f_t[\mathbf{s}(\mathbf{q};t),t]=\frac{df[\mathbf{s}(\mathbf{q};t),t]}{dt}:=\lim_{dt\to
	     0}\frac{f[\mathbf{s}(\mathbf{q};t),t+dt]-f[\mathbf{s}(\mathbf{q};t),t]}{dt}=f_t[\mathbf{q},t;\mathfrak{S}(\mathbf{q};t)]~.
	 \end{align*}
	 The indefinite integral of $f_t$ along time is
	 \begin{align*}
	   f[\mathbf{s}(\mathbf{q};t),t]=\int_{\theta}^t
	   f_t[\mathbf{q},u;\mathfrak{S}(\mathbf{q};t)]du \quad ,
	   \quad \theta\in[0,t]~.
	 \end{align*}
	 
	 \hspace{3mm}
	 Starting from the fact that $s_i(\mathbf{q};t)$ are
	 independent variables in $\mathbf{s}$ and following the way
	 usually employed in vector analysis, see e.g. \cite{Spain},
	 the expressions for the basic differential operators in
	 $\mathbf{s}$ (see the appendix) are 
	 \begin{align*}
	   &\boldsymbol{\nabla}:= \partial_{s_i}\mathbf{I}_i  \quad ,\quad \nabla\cdot
	   := {}^i\partial_{s_i} \qquad , \quad \boldsymbol{\nabla}\times
	   :=[^{m(i+2)}\partial_{s_{m(i+1)}}-^{m(i+1)}\partial_{s_{m(i+2)}}]\mathbf{I}_i~,\\
	   &\nabla^2:= \partial_{s_i^2} \quad~,\quad \boldsymbol{\nabla}^2 
	   := {}^i\nabla^2\mathbf{I}_i \quad ,\quad \nabla^4
	   :=\partial_{s_i^4}+\partial_{s_i^2s_{m(i+1)}^2}+\partial_{s_i^2s_{m(i+2)}^2}
	   \quad,\quad \boldsymbol{\nabla}^4 :={}^i\nabla^4\mathbf{I}_i~;
	 \end{align*}
	 $\partial$ denotes partial differentiation, a left
	 superscript of an operator indicates the vector component on
	 which it operates and
	 \begin{align*}
	   m(i):=\left\{
	   \begin{array}{rl}
	     &i \mapsto i \qquad~~ , \quad i\in[1,3] \\
	     &i \mapsto i-3 \quad , \quad i\in[4,5]
	   \end{array}\right. .
	 \end{align*}
	 
	 \hspace{3mm}
	 The differential operators in $\mathbf{q}$ are obtained by
	 applying $\mathscr{M}$ on the corresponding operators in
	 $\mathbf{s}$. The operators in
	 $\mathbf{s}$,$\mathbf{q}\in\mathbb{R}^2$ are obtained by
	 those in $\mathbf{s}$,$\mathbf{q}\in\mathbb{R}^3$ by setting
	 $\partial_{s_3}=0$, $\partial_{q_3}=0$.

	 \section{A normalisation of the Navier-Stokes equations in
	   orthogonal plane space}
	 \label{Section_3}
	 
	 \hspace{3mm}Consider the mechanics of fluid of density $\rho$ and
	 kinematic viscosity $\nu$, in plane space $\mathbf{s}\in
	 \mathbb{R}^2$ and in time. Applying a mapping
	 $\mathscr{M}:\{\mathbf{s}\mapsto \mathbf{q}\}$, such as that
	 obtained by a conformal mapping of the cartesian plane in
	 $\mathbf{q}$, it is

	 \begin{align*}
	   \mathfrak{S}(\mathbf{q};t)=\mathfrak{s}_i(\mathbf{q};t)\mathbf{I}_i=\mathfrak{s}(\mathbf{q};t)\mathbf{I}_i ~.
	 \end{align*}
	 
	 \hspace{3mm}
	 The fundamental physical measures, of fluid mechanics in the
	 plane, are the streamfunction
	 $\mathbf{\Psi}=\Psi\mathbf{I}_3$, the velocity
	 $\mathbf{U}=U_i\mathbf{I}_i$, $i=1,2$, the vorticity
	 $\mathbf{\omega}=\omega\mathbf{I}_3$ and the pressure
	 $P$. The Navier-Stokes equations in the plane are
	 \begin{align*}
	   &
	   D_t\mathbf{U}=-\rho^{-1}\boldsymbol{\nabla}P+\nu\boldsymbol{\nabla}^2\mathbf{U}
	   \quad~ , \quad \nabla\cdot \mathbf{U}=0 ~~~~ ,~ \textrm{or}\\
	   &D_t\omega=\nu\nabla^2\omega \qquad \qquad \qquad \quad, \quad \nabla^2\Psi=-\omega ~~~;
	 \end{align*}
	 $D_t$ signifies the material derivative
	 \begin{align*}
	   D_t:=\partial_t+\mathbf{U}\cdot\boldsymbol{\nabla} ~.
	 \end{align*}
	 The kinematic relation between the velocity and the
	 streamfunction for plane motion is
	 \begin{align*}
	   \mathbf{U}=\boldsymbol{\nabla}\times \mathbf{\Psi} ~.
	 \end{align*}
	 The physical measures are functions of the independent
	 variables and the parameters as
	 \begin{align*}
	   &\Psi[\mathbf{s}(\mathbf{q};t),t;\nu;\rho]=\Psi[\mathbf{q},t;\mathfrak{s}(\mathbf{q};t);\nu
	     ;\rho]\quad ,\quad
	   U_i[\mathbf{s}(\mathbf{q};t),t;\nu;\rho]=U_i[\mathbf{q},t;\mathfrak{s}(\mathbf{q};t);\nu
	     ;\rho]~,\\
	   &\omega[\mathbf{s}(\mathbf{q};t),t;\nu;\rho]=\omega[\mathbf{q},t;\mathfrak{s}(\mathbf{q};t);\nu;\rho]\quad
	   ,\quad
	   P[\mathbf{s}(\mathbf{q};t),t;\nu;\rho]=P[\mathbf{q},t;\mathfrak{s}(\mathbf{q};t);\nu;\rho]~.
	 \end{align*}
	 
	 \hspace{3mm}
	 Consider the boundary $\partial \mathbf{q}$ of $\mathbf{q}$
	 to be rectangular and symmetrical with respect to
	 $\mathbf{q}=0$. Set the lines $q_i=\pm l_i$, to confine
	 $\mathbf{q}$ and define $\mathbf{l}=l_i\mathbf{I}_i$. Introduce
	 a characteristic speed $U_s$, a characteristic, cartesian
	 length scale $\hat{l}$ and for $\mathfrak{s}(\mathbf{q};t)$
	 with units of length, set $\mathfrak{s}(\mathbf{q};t)$ to
	 scale length in $\mathbf{s}$ and
	 $\mathfrak{s}(\mathbf{q};t)U_s^{-1}$ time. The characteristic
	 length and time scales, are functions of space and time, but
	 based on the analysis presented in section \ref{Section_2},
	 they are parameters in the differential calculus of the
	 physical measures of fluid motion in $\mathbf{q}$. The scaled
	 variables, parameters and measures are
	 \begin{align*}
	   &\tilde{q}_i :=q_i l_i^{-1} & ,
	   &\quad \tilde{\mathfrak{s}}(\tilde{\mathbf{q}};\tilde{t}):=\mathfrak{s}\hat{l}^{-1}
	   &,
	   & \quad \hat{s}_i(\tilde{\mathbf{q}};\tilde{t}):=s_i(\tilde{\mathbf{q}};\tilde{t})\hat{l}^{-1}
	   & ,\\
	   &\tilde{s}_i(\tilde{\mathbf{q}};\tilde{t};\tilde{\mathfrak{s}}) :=\hat{s}_i(\tilde{\mathbf{q}};\tilde{t})\tilde{\mathfrak{s}}^{-1} & ,
	   & \quad d\hat{s}_i:=l_i\tilde{\mathfrak{s}}d\tilde{q}_i
	   & ,
	   & \quad d\tilde{s}_i:=l_id\tilde{q}_i
	   & ,\\
	   &\tilde{t} :=t U_s\mathfrak{s}^{-1} & ,
	   & \quad Re_{\mathfrak{s}}:=U_s\mathfrak{s}\nu^{-1}
	   & ,
	   & \quad Re_{\hat{l}}:=U_s\hat{l}\nu^{-1}
	   & ,\\
	   &\tilde{\Psi}:=\Psi(U_s\mathfrak{s})^{-1} & ,
	   & \quad \tilde{\mathbf{U}}:=\mathbf{U}U_s^{-1}
	   &,
	   & \quad \tilde{\omega}:=\omega\mathfrak{s}U_s^{-1}
	   & ,\\
	   &\tilde{P}:=P(\rho U_s^2)^{-1} & ,
	   & \quad \hat{\Psi}:=\tilde{\Psi}\tilde{\mathfrak{s}}
	   & ,
	   & \quad \hat{\omega}:=\tilde{\omega}\tilde{\mathfrak{s}}^{-1}
	   \quad .
	 \end{align*}	 
	 Introduce the scaled differential operators
	 \begin{align*}
	   &\tilde{\boldsymbol{\nabla}}:=\mathfrak{s}\boldsymbol{\nabla} \quad
	   &, & \quad
	   \tilde{\nabla}\cdot := \mathfrak{s} \nabla \cdot \quad &,& \quad
	   \tilde{\boldsymbol{\nabla}}\times:=\mathfrak{s}\boldsymbol{\nabla}
	   \times ~&,\\
	   &\tilde{\boldsymbol{\nabla}}^2:=\mathfrak{s}^2\boldsymbol{\nabla}^2
	   \quad &, & \quad
	   D_{\tilde{t}}:=\partial_{\tilde{t}}+\tilde{\mathbf{U}}\cdot\tilde{\boldsymbol{\nabla}}
	   \quad &, & \quad
	   \tilde{\boldsymbol{\nabla}}^4:=\mathfrak{s}^4\mathbf{\boldsymbol{\nabla}}^4 ~ &,
	 \end{align*}
	 and the scaled Navier-Stokes equations in the plane are
	 \begin{align*}
	   &D_{\tilde{t}}\tilde{\mathbf{U}}=-\tilde{\boldsymbol{\nabla}}\tilde{P}+Re_{\mathfrak{s}}^{-1}\tilde{\boldsymbol{\nabla}}^2\tilde{\mathbf{U}}
	   \quad,& \tilde{\nabla}\cdot\tilde{\mathbf{U}}=0 \quad ,\\
	   &D_{\tilde{t}}\tilde{\omega}=Re_{\mathfrak{s}}^{-1}\tilde{\nabla}^2\tilde{\omega}
	   \qquad \qquad \quad~ ,& \tilde{\nabla}^2\tilde{\Psi}=-\tilde{\omega} ~~ ,\\
	   &\tilde{\mathbf{U}}=\tilde{\boldsymbol{\nabla}}\times\tilde{\mathbf{\Psi}}~.
	 \end{align*}
	 The differential equation for $\tilde{\Psi}$ is
	 \begin{align*}
	   D_{\tilde{t}}(\tilde{\nabla}^2\tilde{\Psi})=Re_{\mathfrak{s}}^{-1}\tilde{\nabla}^4\tilde{\Psi}
	 \end{align*}
	 and in explicit form
	 \begin{align*}
	   \tilde{\Psi}_{\tilde{s}_1^4}+2\tilde{\Psi}_{\tilde{s}_1^2\tilde{s}_2^2}+\tilde{\Psi}_{\tilde{s}_2^4}=\left( 
	   \tilde{\Psi}_{\tilde{s}_1^2\tilde{t}}+\tilde{\Psi}_{\tilde{s}_2^2\tilde{t}}+\tilde{\Psi}_{\tilde{s}_2}\tilde{\Psi}_{\tilde{s}_1^3}+
	   \tilde{\Psi}_{\tilde{s}_2}\tilde{\Psi}_{\tilde{s}_1\tilde{s}_2^2}-\tilde{\Psi}_{\tilde{s}_1}\tilde{\Psi}_{\tilde{s}_2\tilde{s}_1^2}-
	   \tilde{\Psi}_{\tilde{s}_1}\tilde{\Psi}_{\tilde{s}_2^3}
	   \right)Re_{\mathfrak{s}} ~ .
	 \end{align*}
	 The functional dependence of the scaled physical measures on
	 the scaled variables and the parameters is
	 \begin{align*}
	   &\tilde{\Psi}(\tilde{\mathbf{s}},\tilde{t};\tilde{\mathfrak{s}};Re_{\mathfrak{s}})=
	   \tilde{\Psi}(\tilde{\mathbf{q}},\tilde{t};\tilde{\mathfrak{s}};Re_{\mathfrak{s}};\mathbf{l})
	   & ,& \quad \tilde{\mathbf{U}}(\tilde{\mathbf{s}},\tilde{t};\tilde{\mathfrak{s}};Re_{\mathfrak{s}})=
	   \tilde{\mathbf{U}}(\tilde{\mathbf{q}},\tilde{t};\tilde{\mathfrak{s}};Re_{\mathfrak{s}};\mathbf{l})
	   ~ , \\
	   &\tilde{\omega}(\tilde{\mathbf{s}},\tilde{t};\tilde{\mathfrak{s}};Re_{\mathfrak{s}})=
	   \tilde{\omega}(\tilde{\mathbf{q}},\tilde{t};\tilde{\mathfrak{s}};Re_{\mathfrak{s}};\mathbf{l})
	   & ,& \quad \tilde{P}(\tilde{\mathbf{s}},\tilde{t};\tilde{\mathfrak{s}};Re_{\mathfrak{s}})=
	   \tilde{P}(\tilde{\mathbf{q}},\tilde{t};\tilde{\mathfrak{s}};Re_{\mathfrak{s}};\mathbf{l})
	   ~ . \\
	 \end{align*}
	 \section{The kinematics of the flow of a fluid around two
	   circular cylinders moving towards central impact}
	 \label{Section_4}
	 
\begin{figure}
	   \includegraphics [scale=0.9]{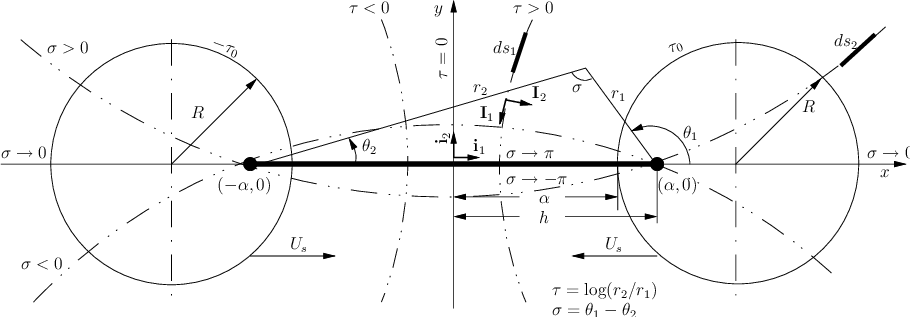}
	   \caption{Description of the plane space between two
	   circular cylinders of zero roughness surfaces according to
	   the bipolar transformation(see also \cite{Milne-Thomson}).}
	   \label{figure_1}
	 \end{figure}

	 \hspace{3mm}Consider two identical, circular cylinders of radius
	 $\hat{l}=R$, immersed in incompressible fluid. The fluid is
	 set into motion due to the motion of the cylinders, with
	 constant and equal in magnitude, but opposite velocity $U_s$
	 towards central impact (see figure
	 \ref{figure_1}). $2h(t)=2h_0-2U_st$ denotes the minimum
	 distance between the boundaries of the cylinders, $h_0$
	 denotes half the initial distance between the cylinders and $[\mp a(t),0]$ are the
	 foci of the bipolar transformation (see
	 e.g. \cite{Milne-Thomson}). $\mathbf{i}_i$ denote the unit
	 vectors of the cartesian plane $(x,y)$. The scaled parameters
	 and space variables in figure \ref{figure_1} are
	 \begin{align*}
	   &\epsilon(t)=hR^{-1} \quad ,\quad
	   \tau_0'(t):=\tau_0(t)\pi^{-1}=\pi^{-1}
	   \sinh^{-1}(aR^{-1})=\pi^{-1} \sinh^{-1}
	   \left\{[\epsilon(2+\epsilon)]^{1/2}\right\}\quad &,\\
	   &\tilde{\tau}(\tilde{x},\tilde{y};\tau_0'):=\tau(\pi\tau_0')^{-1}=(\pi\tau_0')^{-1}\tanh^{-1}\left\{ 
	   2\sinh(\pi\tau_0')\tilde{x}[\tilde{x}^2+\tilde{y}^2+\sinh^2(\pi\tau_0')]^{-1}
	   \right\} \quad &,\\
	   &
	   \tilde{\sigma}(\tilde{x},\tilde{y};\tau_0'):=\sigma\pi^{-1}=\pi^{-1}\tan^{-1}\left\{ 
	   2\sinh(\pi\tau_0')\tilde{y}[\tilde{x}^2+\tilde{y}^2-\sinh^2(\pi\tau_0')]^{-1}
	   \right\} \quad &,\\
	   &\tilde{x}(\tilde{\sigma},\tilde{\tau};\tau_0'):=xR^{-1}=\sinh(\pi\tau_0')\sinh(\pi\tau_0'\tilde{\tau})
	   [\cosh(\pi\tau_0'\tilde{\tau})-\cos(\pi\tilde{\sigma})]^{-1} &,\\
	   &\tilde{y}(\tilde{\sigma},\tilde{\tau};\tau_0'):=
	   yR^{-1}=\sinh(\pi\tau_0')\sin(\pi\tilde{\sigma})[\cosh(\pi\tau_0'\tilde{\tau})-\cos(\pi\tilde{\sigma})]^{-1}&,\\
	   &l_i\tilde{q}_i\mathbf{I}_i:=\pi\tilde{\sigma}\mathbf{I}_1+\pi\tau_0'\tilde{\tau}\mathbf{I_2}~,
	   \quad
	   \tilde{\mathfrak{s}}(\tilde{\sigma},\tilde{\tau};\tau_0')=\sinh(\pi\tau_0')
	   [\cosh(\pi\tau_0'\tilde{\tau})-\cos(\pi\tilde{\sigma})]^{-1}&, \\
	   &\hat{\mathbf{s}}(\tilde{\sigma},\tilde{\tau};\tau_0')=\pi\int_0^{\tilde{\sigma}}\tilde{\mathfrak{s}}(u,\tilde{\tau};\tau_0')du\mathbf{I}_1+
	   \pi\tau_0'\int_0^{\tilde{\tau}}\tilde{\mathfrak{s}}(\tilde{\sigma},u;\tau_0')du \mathbf{I}_2~.
	 \end{align*}
	 The differential equation for $\tilde{\Psi}$, to solve in
	 $\tilde{\sigma}\in[-1,+1]$, $\tilde{\tau}\in[-1,+1]$, $t\ge
	 0$, is
	 \begin{align*}
	   &\tilde{\Psi}_{\tilde{\tau}^4}+2\tilde{\Psi}_{\tilde{\tau}^2\tilde{\sigma}^2}{\tau_0'}^2+\tilde{\Psi}_{\tilde{\sigma}^4}{\tau_0'}^4=
	   [(\tilde{\Psi}_{\tilde{\tau}^2\tilde{t}}+\tilde{\Psi}_{\tilde{\sigma}^2\tilde{t}}{\tau_0'}^2)\pi^2\tau_0'+\\
	   &+\tilde{\Psi}_{\tilde{\tau}}\tilde{\Psi}_{\tilde{\sigma}\tilde{\tau}^2}-
	     \tilde{\Psi}_{\tilde{\sigma}}\tilde{\Psi}_{\tilde{\tau}^3}+\tilde{\Psi}_{\tilde{\tau}}\tilde{\Psi}_{\tilde{\sigma}^3}{\tau_0'}^2
	     -\tilde{\Psi}_{\tilde{\sigma}}\tilde{\Psi}_{\tilde{\tau}\tilde{\sigma}^2}{\tau_0'}^2](\tau_0'Re_{\mathfrak{s}})~.
	 \end{align*}
	 The functional dependence of the physical measures, on the
	 scaled variables and the parameters, in $\mathbf{q}$, is
	 \begin{align*}
	   \tilde{\Psi}(\tilde{\sigma},\tilde{\tau},\tilde{t};\tilde{\mathfrak{s}};Re_{\mathfrak{s}};\tau_0')~,~
	   \tilde{\mathbf{U}}(\tilde{\sigma},\tilde{\tau},\tilde{t};\tilde{\mathfrak{s}};Re_{\mathfrak{s}};\tau_0')~,~
	   \tilde{\omega}(\tilde{\sigma},\tilde{\tau},\tilde{t};\tilde{\mathfrak{s}};Re_{\mathfrak{s}};\tau_0')~,~
	   \tilde{P}(\tilde{\sigma},\tilde{\tau},\tilde{t};\tilde{\mathfrak{s}};Re_{\mathfrak{s}};\tau_0')~.
	 \end{align*}
	 The boundary and initial conditions for the streamfunction
	 are
	 \begin{align*}
	   &\tilde{\Psi}(\tilde{\sigma},\pm
	   1,\tilde{t};\tilde{\mathfrak{s}};Re_{\mathfrak{s}};\tau_0')=\mp
	   V(\tilde{\sigma};\tilde{\mathfrak{s}};\tau_0') &,&\quad
	   \tilde{\Psi}_{\tilde{\tau}}(\tilde{\sigma},\pm
	   1,\tilde{t};\tilde{\mathfrak{s}};Re_{\mathfrak{s}};\tau_0')=0
	   &,\\
	   &\tilde{\Psi}(0,0,\tilde{t};\tilde{\mathfrak{s}};Re_{\mathfrak{s}};\tau_0')=0
	   & , &\quad \tilde{\Psi}(\tilde{\sigma},\tilde{\tau},0;\tilde{\mathfrak{s}};Re_{\mathfrak{s}};\tau_0')=
	   \tilde{\psi}(\tilde{\sigma},\tilde{\tau};\tilde{\mathfrak{s}};Re_{\mathfrak{s}};\tau_0')
	   &, 
	 \end{align*}
	 where
	 $V(\tilde{\sigma};\tilde{\mathfrak{s}};\tau_0')=\tilde{y}(\tilde{\sigma},\pm
	 1;\tau_0')\tilde{\mathfrak{s}}^{-1}=\tilde{y}'(\tilde{\sigma};\tau_0')\tilde{\mathfrak{s}}^{-1}$
	 and
	 $\tilde{\psi}(\tilde{\sigma},\tilde{\tau};\tilde{\mathfrak{s}};Re_{\mathfrak{s}};\tau_0')$
	 denotes the initial flow field.
	 
	 \hspace{3mm}
	 \emph{Remark on the no-slip boundary condition}: The motion of
	 the cylinders is along the $\mathbf{i}_1$-direction. Therefore, the
	 no-slip boundary condition is
	 \begin{align*}
	   \tilde{\Psi}_{\tilde{x}}(\tilde{\sigma},\pm 1,\tilde{t},\tilde{\mathfrak{s}};Re_{\mathfrak{s}};\tau_0')=0~.
	 \end{align*}
	 But, $\tilde{\sigma}=\tilde{\sigma}(\tilde{y}';\tau_0')$ and
	 $\tau_{0\tilde{x}}'=0$, therefore,
	 \begin{align*}
	   \tilde{\Psi}_{\tilde{x}}\left[
	   \tilde{\sigma}(\tilde{y}';\tau_0'),\tilde{\tau},\tilde{t};\tilde{\mathfrak{s}};Re_{\mathfrak{s}};\tau_0'\right]=
	   \tilde{\Psi}_{\tilde{\sigma}}\tilde{\sigma}_{\tilde{y}'}\tilde{y}_{\tilde{x}}'+\tilde{\Psi}_{\tilde{\tau}}\tilde{\tau}_{\tilde{x}}~.
	 \end{align*}
	 For arbitrary values of $\tilde{\tau}$,
	 $\tilde{y}'_{\tilde{x}}\neq 0$. However, on the boundaries of
	 the cylinders, it is 
	 \begin{align*}
	   \frac{d\tilde{y}'(\tilde{\sigma};\tau_0')}{d\tilde{x}(\tilde{\sigma},\pm
	     1;\tau_0')}=\frac{d\tilde{y}(\tilde{\sigma},\pm
	     1;\tau_0')}{d \tilde{x}(\tilde{\sigma},\pm 1;\tau_0')}=0~.
	 \end{align*}
	 \hspace{3mm}
	 Therefore, the no-slip boundary condition is imposed by the
	 requirement, that
	 \begin{align*}
	   \tilde{\Psi}_{\tilde{\tau}}(\tilde{\sigma},\pm 1,
	   \tilde{t};\tilde{\mathfrak{s}},Re_{\mathfrak{s}};\tau_0')=0 ~.
	 \end{align*}
	 
	 \hspace{3mm}
	 An asymptotic series solution for the streamfunction, valid
	 for $0<{\tau_0'}^2\ll 1$ and $0<{\tau_0'}^2Re_R\ll 1$ is obtained
	 by expressing the streamfunction and its boundary and initial
	 conditions as 
	 \begin{align*}
	   &\tilde{\Psi}(\tilde{\sigma},\tilde{\tau},\tilde{t};\tilde{\mathfrak{s}};Re_{\mathfrak{s}};\tau_0')=
	   {}_i^j\tilde{\Psi}(\tilde{\sigma},\tilde{\tau},\tilde{t};\tilde{\mathfrak{s}})(\tau_0'Re_{\mathfrak{s}})^i{\tau_0'}^{2j}
	   & , & \quad i,j\in \mathbb{N}_0 &,\\
	   &\tilde{\psi}(\tilde{\sigma},\tilde{\tau};\tilde{\mathfrak{s}};Re_{\mathfrak{s}};\tau_0')=
	   {}_i^j\tilde{\psi}(\tilde{\sigma},\tilde{\tau};\tilde{\mathfrak{s}})(\tau_0'Re_{\mathfrak{s}})^i{\tau_0'}^{2j}
	   & , & \quad i,j\in \mathbb{N}_0 &,\\
	   &{}_i^j\tilde{\Psi}(\tilde{\sigma},\pm 1,
	   \tilde{t};\tilde{\mathfrak{s}})=\mp V(\tilde{\sigma};\tau_0';\tilde{\mathfrak{s}})\delta_{ij0}
	   & ,& \quad
		 {}_i^j\tilde{\Psi}_{\tilde{\tau}}(\tilde{\sigma},\pm
		 1, \tilde{t};\tilde{\mathfrak{s}})=0 &, \\
		 &{}_i^j\tilde{\Psi}(0,0,\tilde{t};\tilde{\mathfrak{s}})=0
		 & , & \quad
		 {}_i^j\tilde{\Psi}(\tilde{\sigma},\tilde{\tau},0;\tilde{\mathfrak{s}})=
		 {}_i^j\tilde{\psi}(\tilde{\sigma},\tilde{\tau};\tilde{\mathfrak{s}}) &;
	 \end{align*}
	 $\delta$ denotes the Kronecker delta function.
	 
	 \hspace{3mm}
	 The solution for the streamfunction is obtained by
	 substituting the assumed asymptotic series in its
	 differential equation and solving according to regular
	 perturbation theory. The velocity and the vorticity are
	 obtained similarly by the kinematic relations between the
	 velocity or the vorticity and the streamfunction.
	 \begin{align*}
	   &\hat{\Psi}(\tilde{\sigma},\tilde{\tau};Re_R;\tau_0')=\left(
	   \frac{1}{2}\tilde{\tau}^3-\frac{3}{2}\tilde{\tau}\right)\tilde{y}'+3
	   \frac{3!}{7!}(\tilde{\tau}^7-3\tilde{\tau}^3+2\tilde{\tau})\tilde{y}'\tilde{y}_{\tilde{\sigma}}'(\tau_0'Re_R)+\\
	   &+O({\tau_0'}^2Re_R^2)+\left[
	   -\frac{3!}{5!}(\tilde{\tau}^5-2\tilde{\tau}^3+\tilde{\tau})\tilde{y}_{\tilde{\sigma}^2}'+O(\tau_0'Re_R)\right]{\tau_0'}^2+
	   O({\tau_0'}^4)\quad ;
	 \end{align*}
	 
	 \begin{align*}
	   &\tilde{\mathbf{U}}(\tilde{\sigma},\tilde{\tau};\tilde{\mathfrak{s}};Re_{\mathfrak{s}};\tau_0')=\pi^{-1}\left[
	     {}_i^j\tilde{\Psi}_{\tilde{\tau}}(\tilde{\sigma},\tilde{\tau};\tilde{\mathfrak{s}}){\tau_0'}^{-1}\mathbf{I}_1-
	     {}_i^j\tilde{\Psi}_{\tilde{\sigma}}(\tilde{\sigma},\tilde{\tau};\tilde{\mathfrak{s}})\mathbf{I}_2\right](\tau_0'Re_{\mathfrak{s}})^i{\tau_0'}^{2j}~,\\
	   &\tilde{\mathbf{U}}
	   (\tilde{\sigma},\tilde{\tau};\tilde{\mathfrak{s}};Re_R;\tau_0')=\left\{
	   \left[\frac{3}{2}\left(\tilde{\tau}^2-1
	     \right)\tilde{y}'\tilde{\mathfrak{s}}^{-1}+3\frac{3!}{7!}\left(
	     7\tilde{\tau}^6-9\tilde{\tau}^2+2\right)\tilde{y}'\tilde{y}'_{\tilde{\sigma}}\tilde{\mathfrak{s}}^{-1}(\tau_0'Re_R)+\right.\right.\\
	     &\left.\left.+O({\tau_0'}^2Re_R^2)\right]\pi^{-1}
		{\tau_0'}^{-1}+O(\tau_0')\right\}\mathbf{I}_1-\left[\left(
	     \frac{1}{2}\tilde{\tau}^3-\frac{3}{2}\tilde{\tau}\right)\tilde{y}'_{\tilde{\sigma}}\tilde{\mathfrak{s}}^{-1}\pi^{-1}+\right.\\
	     &\left.+3\frac{3!}{7!} (\tilde{\tau}^7-3\tilde{\tau}^3+2\tilde{\tau})[(\tilde{y}'_{\tilde{\sigma}})^2+
	       \tilde{y}'\tilde{y}'_{\tilde{\sigma}^2}]\tilde{\mathfrak{s}}^{-1}\pi^{-1}(\tau_0'Re_R)+O({\tau_0'}^2Re_R^2)+O({\tau_0'}^2)\right]\mathbf{I}_2~;
	 \end{align*}
	 
	 \begin{align*}
	   &\tilde{\omega}(\tilde{\sigma},\tilde{\tau};\tilde{\mathfrak{s}};Re_{\mathfrak{s}};\tau_0')=
	   -\pi^{-2}\left[{}_i^j\tilde{\Psi}_{\tilde{\tau}^2}(\tilde{\sigma},\tilde{\tau};\tilde{\mathfrak{s}}){\tau_0'}^{-2}+{}_i^j\tilde{\Psi}_{\tilde{\sigma}^2}(
	     \tilde{\sigma},\tilde{\tau};\tilde{\mathfrak{s}})
	     \right](\tau_0'Re_{\mathfrak{s}})^i{\tau_0'}^{2j}~,\\
	   &\hat{\omega}(\tilde{\sigma},\tilde{\tau};\tilde{\mathfrak{s}};Re_R;\tau_0')=-\left[3\tilde{\tau}\tilde{y}'\tilde{\mathfrak{s}}^{-2}+
	     3\frac{3!^2}{7!}(7\tilde{\tau}^5-3\tilde{\tau})\tilde{y}'\tilde{y}'_{\tilde{\sigma}}\tilde{\mathfrak{s}}^{-2}(\tau_0'Re_R)+\right.\\
	     &+\left.O({\tau_0'}^2Re_R^2)\right](\pi^2{\tau_0'}^2)^{-1}+O(1)~.
	 \end{align*}
	 
	 \hspace{3mm}
	 For $0<{\tau_0'}^2\ll 1$ and $0<{\tau_0'}^2Re_R\ll 1$, when
	 the asymptotic series solution is valid, the solution is
	 independent of the initial condition and unique.
	 
	 \hspace{3mm}
	 For comparison, consider the case of irrotational fluid
	 motion. For this case, the equation and the boundary
	 conditions for the streamfunction
	 $\tilde{\Psi}(\tilde{\sigma},\tilde{\tau};\tilde{\mathfrak{s}};\tau_0')$,
	 are
	 \begin{align*}
	   &{\tau_0'}^2\tilde{\nabla}^2\tilde{\Psi}=\tilde{\Psi}_{\tilde{\tau}^2}+{\tau_0'}^2\tilde{\Psi}_{\tilde{\sigma}^2}=0\quad
	   ,\\
	   &\tilde{\Psi}(\tilde{\sigma},\pm
	   1;\tilde{\mathfrak{s}};\tau_0')=\mp
	   V(\tilde{\sigma};\tilde{\mathfrak{s}};\tau_0') \quad, \quad
	   \tilde{\Psi}(0,0;\tilde{\mathfrak{s}};\tau_0')=0 ~.
	 \end{align*}
	 An asymptotic series solution for the streamfunction, for
	 irrotational flow, valid for $0<{\tau_0'}^2\ll 1$, is
	 obtained by expressing the streamfunction and its boundary
	 conditions as
	 \begin{align*}
	   &\tilde{\Psi}(\tilde{\sigma},\tilde{\tau};\tilde{\mathfrak{s}};\tau_0')={}^j\tilde{\Psi}(\tilde{\sigma},\tilde{\tau};\tilde{\mathfrak{s}}){\tau_0'}^{2j}~,
	   \quad j\in \mathbb{N}_0~,\\
	   &{}^j\tilde{\Psi}(\tilde{\sigma},\pm
	   1;\tilde{\mathfrak{s}})=\mp
	   V(\tilde{\sigma};\tilde{\mathfrak{s}};\tau_0')\delta_{j0}~,
	   \quad {}^j\tilde{\Psi}(0,0;\tilde{\mathfrak{s}})=0 \quad .
	 \end{align*}
	 The solution for the streamfunction, for irrotational flow,
	 is obtained by substituting the assumed asymptotic series in
	 the Laplace equation for the streamfunction and solving
	 according to regular perturbation theory to obtain
	 \begin{align*}
	   \tilde{\Psi}(\tilde{\sigma},\tilde{\tau};\tilde{\mathfrak{s}};\tau_0')=(-1)^{j+1}G(\tilde{\tau};2j)V_{\tilde{\sigma}^{2j}}~{\tau_0'}^{2j}~,
	   \qquad j\in\mathbb{N}_0~, \textrm{where}
	 \end{align*}
	 $G(\tilde{\tau};2j)$ denote the polynomial functions of
	 $\tilde{\tau}$, expressed recursively as
	 \begin{align*}
	   &G(\tilde{\tau};0):=\tilde{\tau} \quad ; \quad
	   G(\tilde{\tau};2j-1):=\frac{\tilde{\tau}^{2j+1}-\tilde{\tau}}{(2j+1)!}
	   \quad ,  \forall j\in \mathbb{N}_1 \quad ; \quad
	   G(\tilde{\tau};2):=G(\tilde{\tau};1)~;\\
	   &G(\tilde{\tau};2j):=G(\tilde{\tau};2j-1)-\frac{G(\tilde{\tau};2k+2)}{[2(j-k)-1]!}\qquad,
	   ~ k\in\mathbb{N}_0:k\in[0,j-2]\quad ,\\
	   &\forall j\in\mathbb{N}: j\in[2,+\infty)~.
	 \end{align*}
	 The velocity vector, for irrotational flow is then
	 \begin{align*}
	   \tilde{\mathbf{U}}(\tilde{\sigma},\tilde{\tau};\tilde{\mathfrak{s}};\tau_0')=(-1)^{j+1}G_{\tilde{\tau}}(\tilde{\tau};2j)V_{\tilde{\sigma}^{2j}}\pi^{-1}
	   {\tau_0'}^{2j-1}\mathbf{I}_1+(-1)^jG(\tilde{\tau};2j)V_{\tilde{\sigma}^{2j+1}}\pi^{-1}{\tau_0'}^{2j}\mathbf{I}_2~.
	 \end{align*}
	 
	 \hspace{3mm}
	 \emph{Remark on the unsteady fluid motion:} The flow field,
	 induced by the motion of the circular cylinders, is time
	 dependent. The present formulation separates the effects of
	 the unsteady fluid motion into two parts; a) fluid motion
	 induced solely by and which is synchronised with the motion of
	 the boundaries of the flow field and b) time-dependent fluid
	 motion, which is independent of the motion of the
	 boundaries. For irrotational flow, the second kind of fluid
	 motion can be induced only by imposing time-dependent
	 kinematic boundary conditions. For rotational flow,
	 however, inertial effects, which are described by the
	 $\partial_{\tilde{t}}$ derivative, can also induce unsteady
	 fluid motion of the second kind.
	 
	 \hspace{3mm}
	 For the present case of fluid motion, induced by the motion
	 of the circular cylinders with constant velocity towards
	 impact, the unsteady fluid motion is only of the first kind
	 at the asymptotic limit of approach of the cylinders. For
	 irrotational flow, this is only because the kinematic
	 boundary conditions are independent of time.  For rotational
	 flow, this is because of the steady boundary conditions
	 and the fact that the $\partial_{\tilde{t}}$ derivative does
	 not appear in the differential equation for the leading order
	 term of the asymptotic series for the streamfunction. The
	 implication of this, is that the initial condition becomes
	 redundant at the asymptotic limit considered and that the
	 asymptotic series solution is independent of any initial flow
	 field and thus, unique.

	 \section{An interpretation of the asymptotic series solution}
	 \label{Section_5}
	 
	 \hspace{3mm}Consider the asymptotic series of the streamfunction for
	 rotational and irrotational flow, respectively,
	 \begin{align*}
	   \hat{\Psi}(\tilde{\sigma},\tilde{\tau};Re_R;\tau_0')={}_i^j\hat{\Psi}(\tilde{\sigma},\tilde{\tau})(\tau_0'Re_R)^i{\tau_0'}^{2j}\quad
	   , \quad
	   \hat{\Psi}(\tilde{\sigma},\tilde{\tau};\tau_0')={}^j\hat{\Psi}(\tilde{\sigma},\tilde{\tau}){\tau_0'}^{2j}
	   \quad .
	 \end{align*}
	 The streamfunction for irrotational flow, is decomposed
	 into infinite kinematic scales of volume flow rates of order
	 $O[(U_sR){\tau_0'}^{2j}]$, described by
	 ${}^j\hat{\Psi}(\tilde{\sigma},\tilde{\tau})$~. The
	 streamfunction for rotational flow is also decomposed, into
	 infinite kinematic scales of volume flow rates of order
	 $O[(U_sR){\tau_0'}^{2j}]$. However, for this case, every
	 kinematic scale is decomposed further, into infinite dynamic
	 scales according to the dynamic scaling
	 $(\tau_0'Re_R)^i$. The largest scale of the flow field of
	 order $O(U_sR)$ is described by
	 ${}_0^0\hat{\Psi}(\tilde{\sigma},\tilde{\tau})$. The lower
	 order terms,
	 ${}_i^j\hat{\Psi}(\tilde{\sigma},\tilde{\tau})(1-\delta_{ij0})$,
	 of the asymptotic series solution, are present and non-zero
	 for all $\tau_0'$, but, as $\tau_0'\to 0$, they represent
	 asymptotically vanishing physical flow scales of order
	 $O[U_sR(\tau_0'Re_R)^i{\tau_0'}^{2j}]$.
	 ${}_i^j\hat{\Psi}(\tilde{\sigma},\tilde{\tau})$ are
	 inter-dependent according to the differential equation
	 \begin{align*}
	   \mathscr{D}~[{}_i^j\hat{\Psi}(\tilde{\sigma},\tilde{\tau})](\tau_0'Re_R)^i{\tau_0'}^{2j}=0~,
	 \end{align*}
	 where
	 $\mathscr{D}=(D_{\tilde{t}}\tilde{\nabla}^2-Re_{\mathfrak{s}}^{-1}\tilde{\nabla}^4)$.
	 
	 \hspace{3mm}
	 According to the present interpretation, the initial condition
	 $\hat{\psi}(\tilde{\sigma},\tilde{\tau};Re_R;\tau_0')=$
	 \newline
	 $={}_i^j\hat{\psi} (\tilde{\sigma},\tilde{\tau}) (\tau_0'Re_R)^i{\tau_0'}^{2j}$,
	 for the rotational flow field, in section \ref{Section_4},
	 represents a multi-scale initial flow field. The magnitude of
	 these scales, however, is adjusted according to the change of
	 the $\tau_0'$ parameter, although the flow field at every
	 scale can be arbitrary.
	 
	 \hspace{3mm}
	 At the limit $Re_R\to 0$, the asymptotic series solution, for
	 rotational fluid motion, becomes an asymptotic series
	 solution for $\tilde{\nabla}^4\tilde{\Psi}=0$, as expected.
	 
	 \section{Acknowledgements}
	 
	 The author wishes to thank Professor J.M.R. Graham for introducing him to the problem of impact of two circular cylinders. 
	 He also acknowledges fruitfull discussions on perturbation methods with Dr Mario Sandoval Espinoza, on turbulence with Dr Vassilios Dallas 
	 and on fluid mechanics with Dr Joanna Isabelle Whelan.

	 \appendix
	 \section{Calculation of the basic differential operators in the $s$-space.}
	 \label{Appendix}
	 The gradient $\boldsymbol{\nabla}$ of a scalar $f(\mathbf{s})$ is
	 \begin{align*}
	   \boldsymbol{\nabla}f(\mathbf{s})=f_{s_i}(\mathbf{s})\mathbf{I}_i~.
	 \end{align*}
	 $s_i$ are independent variables; therefore,
	 \begin{align*}
	   \mathbf{I}_i=\mathbf{I}_{m(i+1)}\times\mathbf{I}_{m(i+2)}=\boldsymbol{\nabla}s_i=\boldsymbol{\nabla}s_{m(i+1)}\times\boldsymbol{\nabla}s_{m(i+2)}~.
	 \end{align*}
	 The divergence $\nabla\cdot$ of a vector
	 $\mathbf{f}(\mathbf{s})=f_i(\mathbf{s})\mathbf{I}_i$ is
	 \begin{align*}
	   &\nabla \cdot [f_i(\mathbf{s})\mathbf{I}_i]=\nabla \cdot
	   \left\{f_i(\mathbf{s})\cdot
		   [\boldsymbol{\nabla}s_{m(i+1)}\times \boldsymbol{\nabla}s_{m(i+2)}] \right \}=\\
		   &=f_i(\mathbf{s})\nabla\cdot
		   [\boldsymbol{\nabla}s_{m(i+1)}\times
		     \boldsymbol{\nabla}s_{m(i+2)}]+[\boldsymbol{\nabla}s_{m(i+1)}\times
		     \boldsymbol{\nabla}s_{m(i+2)}]\cdot \boldsymbol{\nabla}
		   f_i(\mathbf{s})=\\
		   &=f_i(\mathbf{s})\{\boldsymbol{\nabla}s_{m(i+2)}\cdot[\boldsymbol{\nabla}\times\boldsymbol
		     {\nabla}s_{m(i+1)}]-\boldsymbol{\nabla}s_{m(i+1)}\cdot[\boldsymbol{\nabla}\times
		     \boldsymbol{\nabla}s_{m(i+2)}] \}+\\
		   &+[\boldsymbol{\nabla}s_{m(i+1)}\times
		     \boldsymbol{\nabla}s_{m(i+2)}]\cdot
		   \boldsymbol{\nabla}f_i(\mathbf{s})=\mathbf{I}_i\cdot \boldsymbol{\nabla}f_i(\mathbf{s})={f_i}_{s_i}(\mathbf{s})~.
	 \end{align*}
	 Therefore,
	 \begin{align*}
	   \nabla\cdot \mathbf{f}(\mathbf{s})=f_{i_{s_i}}(\mathbf{s})~.
	 \end{align*}
	 The local rotation of an arbitrary vector field
	 $\boldsymbol{\nabla}\times \mathbf{f}(\mathbf{s})$ is obtained as
	 \begin{align*}
	   &\boldsymbol{\nabla}\times
	   [f_i(\mathbf{s})\mathbf{I}_i]=\boldsymbol{\nabla}\times
	   [f_i(\mathbf{s})\boldsymbol{\nabla}s_i]=
	   f_i(\mathbf{s})\boldsymbol{\nabla}\times\boldsymbol{\nabla}s_i+\boldsymbol{\nabla}f_i(\mathbf{s})\times
	   \boldsymbol{\nabla}s_i=\\
	   &=\boldsymbol{\nabla}f_i(\mathbf{s})\times \mathbf{I}_i={f_i}_{s_{m(i+2)}}(\mathbf{s})\mathbf{I}_{m(i+1)}-{f_i}_{s_{m(i+1)}}(\mathbf{s})\mathbf{I}_{m(i+2)}~.
	 \end{align*}
	 Therefore, 
	 \begin{align*}
	   \boldsymbol{\nabla}\times \mathbf{f}(\mathbf{s})=\left[ {f_{m(i+2)}}_{s_{m(i+1)}}(\mathbf{s})-{f_{m(i+1)}}_{s_{m(i+2)}}(\mathbf{s})\right]\mathbf{I}_i~.
	 \end{align*}
	 The Laplace operator $\nabla^2$ of a scalar $f(\mathbf{s})$
	 and the operator $\boldsymbol{\nabla}^2$ of a vector
	 $\mathbf{f}(\mathbf{s})$ satisfies the identities
	 \begin{align*}
	   \nabla^2 f(\mathbf{s})=\nabla\cdot
		 [\boldsymbol{\nabla}f (\mathbf{s})] \quad , \quad
		 \boldsymbol{\nabla}^2\mathbf{f} (\mathbf{s})=\boldsymbol{\nabla}
			    [\nabla\cdot\mathbf{f} (\mathbf{s})]-\boldsymbol{\nabla}\times
			    [\boldsymbol{\nabla}\times \mathbf{f}(\mathbf{s})]~.
	 \end{align*}
	 Therefore,
	 \begin{align*}
	   \nabla^2f(\mathbf{s})=f_{s_i^2}(\mathbf{s}) \quad , \quad
	   \boldsymbol{\nabla}^2\mathbf{f}(\mathbf{s})=\nabla^2f_i(\mathbf{s})\mathbf{I}_i~.
	 \end{align*}
	 The biharmonic operator of a scalar $f(\mathbf{s})$ and of a
	 vector $\mathbf{f}(\mathbf{s})$, respectively, is
	 \begin{align*}
	   &\nabla^4f(\mathbf{s})=\nabla^2[\nabla^2f(\mathbf{s})]=f_{s_i^4}(\mathbf{s})+f_{s_i^2s^2_{m(i+1)}}(\mathbf{s})+f_{s_i^2s^2_{m(i+2)}}(\mathbf{s})\quad
	   ,\\
	   &\boldsymbol{\nabla}^4\mathbf{f}(\mathbf{s})=\nabla^4f_i(\mathbf{s})\mathbf{I}_i~.
	 \end{align*}
	 Therefore, in operational form,
	 \begin{align*}
	   &\boldsymbol{\nabla}:=\partial_{s_i}\mathbf{I}_i \quad , \quad
	   \nabla\cdot:={}^i\partial_{s_i} \qquad , \quad
	   \boldsymbol{\nabla}\times:=\left[{}^{m(i+2)}\partial_{s_{m(i+1)}}-{}^{m(i+1)}\partial_{s_{m(i+2)}} \right]\mathbf{I}_i~,\\
	   &\nabla^2:=\partial_{s_i^2} \quad ~, \quad
	   \boldsymbol{\nabla}^2:={}^i\nabla^2\mathbf{I}_i \quad , \quad
	   \nabla^4:=\partial_{s_i^4}+\partial_{s_i^2s^2_{m(i+1)}}+\partial_{s_i^2s^2_{m(i+2)}}
	   \quad , \boldsymbol{\nabla}^4:={}^i\nabla^4\mathbf{I}_i ~.
	 \end{align*}
\end{document}